\documentclass[reprint, twocolumn, aps]{revtex4-2}
\usepackage{amsmath, amssymb}
\usepackage{physics}
\usepackage{graphicx}
\usepackage{hyperref}
\usepackage{color}

\renewcommand\prl{Physical Review Letters}
\renewcommand\nat{Nature}
\renewcommand\apj{The Astrophysical Journal}
\newcommand\araa{Annual Review of Astronomy and Astrophysics}
\newcommand\aap{Advances in Applied Physics}
\newcommand\mnras{MNRAS}
\newcommand\apjl{Astrophysical Journal Letters}
\newcommand\sovast{Soviet Astronomy}
\newcommand\aapr{Astronomy and Astrophysics Reviews}
\newcommand\pasa{Publications of the Astronomical Society of Australia}
\newcommand\aj{Astronomical Journal}
\newcommand\pasj{Publications of the Astronomical Society of Japan}
\newcommand\na{New Astronomy}

\begin{document}


\title{Constraining Dark Matter Inside Stars Using Spectroscopic Binaries\\and a Modified Mass-Luminosity Relation}

\author{Gil Peled}
\affiliation{School of Physics and Astronomy, Tel-Aviv University, Tel-Aviv 69978, Israel}

\author{Tomer Volansky}
\affiliation{School of Physics and Astronomy, Tel-Aviv University, Tel-Aviv 69978, Israel}

\begin{abstract}
The presence of a dissipative dark matter (DM) sector may allow for the trapping of a significant DM mass inside stars, either during structure formation or by accretion over their lifetime, influencing stellar behavior well into the Main Sequence stage. 
Motivated by this scenario, we place an upper bound on the fractional DM mass within current-day Main Sequence stars.
Using double-lined spectroscopic binaries (SB2 stars), gravitational masses are extracted and contrasted with luminous masses, derived using a modified mass-luminosity relation which accounts for the effect of DM.
High-accuracy mass and luminosity data from a sample of 486 binary stars in the $0.18 < M/M_\odot < 31$ mass range are partitioned into appropriate mass domains and analyzed. A 95\% C.L.\ upper limit of sub-5\% is found for the subset of 263 stars in the $1 < M/M_\odot < 2.4$ regime. 
\end{abstract}

\maketitle

\section{Introduction} \label{introduction}
The collisionless cold dark matter (CDM) scenario has been remarkably successful in describing numerous gravitational observations, from the expansion history of our universe to the formation of structures on cosmic and galactic scales. Despite its simplicity, however, recent theoretical considerations suggest a more complex dark matter (DM) which resides in a dark sector and is prone to possible interactions with additional particles and forces (see e.g.~\cite{DarkSector1,DarkSector2,DarkSector3}).  Moreover, various observations on small scales point to such a non-minimal DM scenario~\cite{CDMprob1,CDMprob2,CDMprob3,CDMprob4,CDMprob5,CDMprob6,CDMprob7}, and in particular to a self-interacting dark matter (SIDM) picture, which predicts observable signatures at these scales~\cite{SIDM1,SIDM2,SIDM3,SIDM4,SIDM5,SIDM6,SIDM7,SIDM8,SIDM9,SIDM10,SIDM11}. 

These models, in describing a rich, multi-state dark sector, often include a significant dissipative dark matter (DDM) component, wherein light particles - which possibly mediate the self-interactions - are emitted, causing a dissipation of energy and possibly also angular momentum~\cite{DDMbasics1,DDMbasics2,DDMbasics3,DDMbasics4,DDMbasics5,DDMbasics6}.~While a strongly dissipative component must  necessarily be a small portion of DM~\cite{DDMlimit}, it may lead to interesting measurable effects on structure formation at all scales. Exploration of these effects is underway on the cosmic~\cite{DDMcosmic1,DDMcosmic2,DDMcosmic3}, galactic~\cite{DDMGalact1,DDMGalact2,DDMGalact3,DDMGalact4,DDMGalact5,DDMGalact6,DDMGalact7,DDMGalact8,DDMGalact9,DDMGalact11,DDMGalact12,DDMGalact13,DDMGalact14,DDMGalact15,DDMGalact16,DDMGalact17,DDMGalact18,DDMGalact19} and subgalactic~\cite{DDMSubgalact1,DDMSubgalact2,DDMSubgalact3,DDMSubgalact4,DDMSubgalact5,DDMSubgalact6} scales.  It may prove instructive to extend this exploration, seeking to place constraints on DDM at the stellar scale, where it is yet to be significantly studied (see, however,~\cite{DDMStellar1,DDMStellar2,DDMStellar3}).  This is the main motivation for this work. 

Concurrently, much work has been done on the distribution of DM at the sub-galactic and even local solar neighbourhood scales \cite{DMDist1,DMDist2,DMDist3,DMDist4,DMDist5,DMDist6,DMDist7,DMDist8,DMDist9,DMDist10,DMDist11}. This has been joined by in-depth exploration of the possible means of DM capture or accretion into stars~\cite{StellarDM1,StellarDM2,StellarDM3,StellarDM4,StellarDM5,StellarDM6,StellarDM7,StellarDM8,StellarDM9,StellarDM10,StellarDM11,StellarDM12,StellarDM13,StellarDM14,StellarDM15,StellarDM16,StellarDM17,StellarDM18,StellarDM19,StellarDM20,StellarDM21,StellarDM22,StellarDM23,StellarDM24}, and the impact the presence of DM may have on the star itself, whether by annihilations fuelling specific stages in early stellar evolution in so-called Dark Stars~\cite{DarkStars1,DarkStars2,DarkStars3,DarkStars4,DarkStars5,DarkStars6,DarkStars7,DarkStars8,DarkStars9,DarkStars10,DarkStars11,DarkStars12}, helioseismological effects~\cite{Helioseismo1,Helioseismo2,Helioseismo3,Helioseismo4}, or by axion interactions creating a neutrino signal~\cite{AxionNeutr1,AxionNeutr2,AxionNeutr3} to be detected in indirect detection experiments~\cite{DetectExp1,DetectExp2,DetectExp3}.~However, these studies use traditional non-disspative DM models, in which capture through rare scattering events or Bondi accretion brings a mild DM population into the star, 
and therefore consider fairly small current DM densities. 

In contrast, DDM models may naturally result in significantly larger DM densities in stars, accumulated during initial fragmentation and structure formation or over the stellar lifetime, and remaining in the stellar core indefinitely, well into the Main Sequence stage. In this picture, despite being a small portion of the dark sector, DDM may be the dominant form of DM inside the star. Placing an upper bound on the possible amount of DM present within current-day stars is therefore an important link  in constraining the possible nature of DDM.

In this work we use a simple model for the effect DM may have on stellar luminosity in order to set such an upper bound. The classical mass-luminosity relation (MLR) is re-derived in a modified form to include the presence of DM mass within the stellar core.~This is done using a homologous, or self-similar, stellar model. 
These are models that allow for analytical derivations of relations between stellar parameters, and are useful in interpreting the results of state-of-the-art numerical stellar interior simulations~\cite{MLRDerivHomol1,MLRDerivHomol2,MLRDerivHomol3,MLRDerivHomol4,MLRDerivHomol5,MLRDerivHomol6}.

The resulting DM-modified MLR is then used to compare the gravitational mass and luminosity data available from observations of double-lined spectroscopic binary star systems (SB2s). Photometric and spectrometric measurements of such systems produce very high-accuracy data for stellar parameters with well-defined systematic errors~\cite{SB2Basics1,SB2Basics2,SB2Basics3,SB2Basics4,SB2Basics5,SB2Basics6}, which allow us to extract novel and strong constraints on the DM mass that may hide within stars. 

\section{Mass-Luminosity Relation with a DM Component} \label{Model Derivation}

To place constraints on the DM mass within stars, one must first model the MLR with and without DM.  For this we consider a simple model of a star: spherically-symmetric, in hydrostatic and thermal equilibrium, and with negligible effects from external tidal forces or magnetic fields.   As is known~\cite{MLRDerivHomol1,MLRDerivHomol2} and we summarize below, these assumptions allow for a simple and general analytic MLR, which agrees well with Main Sequence data, used in this work (see Fig.~\ref{fig:Data with segments}).  

In addition, we make two assumptions regarding the effects of DM:
firstly, that DM has negligible effect on the opacity $\kappa$, i.e.~the DM-photon interaction rate is small enough to have little impact on the stellar structure; and secondly, that DM has negligible direct effect on the energy generation per unit mass $\epsilon$.  These assumptions imply that at present times, DM does not play a direct role in energy production, even if it did so in earlier stages (see e.g.~\cite{DarkStars1,DarkStars2,DarkStars3,DarkStars4}).
Therefore $\kappa$ and $\epsilon$ are independent of DM density.

Under these assumptions, the stellar structure can be fully described by a set of 5 equations: hydrostatic equilibrium,  mass continuity, energy transport, thermal equilibrium, and the equation of state.~These are modified from their classical forms by the presence of DM through a \textit{DM quotient factor}, 
\begin{equation}
Q \equiv \frac{M_{\rm tot}}{M_b}\,.
\end{equation}
This definition and the aforementioned stellar structure conditions imply the following set of equations (for a derivation see App.~\ref{appendix:Basic equations}),
\begin{subequations}
\begin{align} 
\pdv{P}{m_b} &= -\frac{G_N\, q(r)\, m_b(r)}{4\pi r^4}\,, \label{eq:basic_HE} \\
\pdv{r}{m_b} &= \frac{1}{4\pi r^2 \rho_b} \,, \label{eq:basic_MCont}\\
\pdv{T}{m_b} &= -\frac{T}{P}\frac{G_N\, q(r)\, m_b(r)}{4\pi r^4} \nabla\,, \label{eq:basic_ET} \\
\pdv{l}{m_b} &= \epsilon \,, \label{eq:basic_TE}\\
P &= \frac{k_B}{m_p\mu}\rho_b T + \frac{1}{3}aT^4  \label{eq:basic_EoS}\,.
\end{align}
\end{subequations}
Here $P$ is the pressure experienced by baryons, $l$ is the local luminosity, $G_N$, $k_B$, and $a$ are the gravitational, Boltzmann, and radiation density constants respectively, $m_p$ is the proton mass, $\mu$ is the mean molecular mass, and $q(r)$ ($m_b(r)$) is the radius-dependent DM quotient factor (baryonic mass).
Moreover (see e.g.~\cite{MLRDerivHomol1}),  
\begin{gather*}
\nabla = \rm min(\nabla_{rad}, \nabla_{ad})\,, \\
\nabla_{\rm rad} = \qty(\pdv{\log T}{\log P})_{\rm rad} =\frac{3}{16\pi a}\frac{\kappa l P}{G_N\, q(r)\, m_b(r)\, T^4}\,, \\
\nabla_{\rm ad} = \qty(\pdv{\log T}{\log P})_{\rm s}\,. 
\end{gather*}

A complete solution for this system of coupled equations is unnecessarily model-dependent when fitting to data.~Instead, a simple scaling relation may be derived by employing a homologous (or ``self-similar") stellar model, which delivers the required MLR functional form to which we fit. 

The derivation of the scaling relation is given in App.~\ref{appendix:Homologous Stellar Models}.  The result strongly depends on the stellar opacity and energy generation. In the simple case of gas-dominated pressure and electron scattering-dominated opacity, $\kappa$ is independent of $\rho_b$ and $T$, which leads directly to the traditional simplified form of the MLR, modified by the DM quotient factor,
\begin{equation}
L \propto Q M_{\rm tot}^3\,.
\end{equation}
Conversely, for radiation-dominated pressure, one finds the luminosity no longer depends on the DM quotient factor.  
More generally, the DM-modified MLR produced by the homologous stellar model takes the form
\begin{equation}
L = \eta \, Q^{\omega}  M_{\rm tot}^\alpha\,,
\label{eq:DM-MLR for Pgas}
\end{equation}
$\eta,\  \omega$ and $\alpha$ being non-negative parameters whose values are determined solely by the Standard Model (SM) physics of the internal stellar structure and composition.  $\omega=0$ corresponds to the case of dominant radiation pressure. Strictly speaking $\omega$ may also significantly exceed 1; however, for the purposes of this work we cap its value at unity, since higher values create degeneracy issues for our statistical analysis. This has the added benefit of ensuring a conservative upper bound for $Q$.

A remark on the validity of the above MLR model is in order.   Na\"ively, Eq.~\eqref{eq:DM-MLR for Pgas} suggests that as $Q$ increases the luminosity can counter-intuitively become arbitrarily large.  
This, however, is wrong for two reasons.  Firstly, as $T\propto Q^{\nu}$ for a positive $\nu$ (see Eq.~\eqref{eq:basic_ET}), the temperature rises together with $Q$, quickly reaching radiation-dominated pressure, for which the luminosity is no longer $Q$-dependent.   Secondly, for $Q$ even slightly above 1, the stellar lifetime on the Main Sequence, which is inversely proportional to the luminosity $\tau_{\rm nuc} \propto M_b/L \propto Q^{-4}M_b^{-2}$~\cite{MLRDerivHomol1,StellarLifetime1}, significantly decreases in a manner inconsistent with observations.  We conclude that Eq.~\eqref{eq:DM-MLR for Pgas} above is valid only for $Q-1 \ll 1$, in agreement with the results derived below.  For further discussion, see App.~\ref{appendix:Homologous Stellar Models}.

\section{The Data Set} \label{Data}
\subsection{The Eker Catalogue}
Our data analysis is based on the 2014 and 2018 catalogues of stellar parameters collected in \cite{EkerData1,EkerData2,EkerData3,EkerData4} which include 586 stars from eclipsing double-lined spectroscopic binary systems (SB2s). This is the largest current collection of high-accuracy mass and luminosity data specifically from SB2s.

The authors filter this sample to remove low accuracy measurements as well as stars that are likely to be outside the Main Sequence, reducing the sample to 509 stars (see~\cite{EkerData3} for details). We 
further discard stars that have no recorded error margins for their luminosity measurements, as these are important for our data analysis, leaving us with a sample of 486 stars in the $0.18 < M_{\rm tot}/M_\odot < 31$ mass range. This final version of the sample is presented in Fig.~\ref{fig:Data with segments}.

The SB2 data allow for the highly accurate extraction of the gravitational mass of each star in the system through radial velocity (RV) curves and the use of the deduced Kepler relation,
\begin{equation}
\frac{M_{1,2}^3 \sin^3 i}{(M_1 + M_2)^2} = \frac{K_{2,1}^3 \tilde P (1-e^2)^\frac{3}{2}}{2\pi G_N}\,.
\end{equation}
Here $i$ is the orbital plane inclination, $\tilde P$ is the orbital period, $e$ is  the eccentricity, and $M_{1,2}$, $K_{1,2}$ are the stellar masses and radial velocity amplitudes respectively, all measured through light curve and spectroscopic RV curve observations (see e.g.~\cite{SB2Basics1,SB2Basics2,SB2Basics3,SB2Basics4,SB2Basics5,SB2Basics6}).

Additionally,~luminosity data is collected either through spectrometry (cf.~e.g.~\cite{EkerSpectroMeas1,EkerSpectroMeas2,EkerSpectroMeas3}), photometry~\cite{EkerPhotoMeas1,EkerPhotoMeas2}, or both~\cite{EkerBothMeas1,EkerBothMeas2,EkerBothMeas3}. Due to this heterogeneity, and due to the range of the data itself, we adopt the common convention of treating the luminosity error distribution as Gaussian in log-space.
As a result, the $\log$ form of Eq.~\eqref{eq:DM-MLR for Pgas} turns out to be more convenient when analyzing the data.

\subsection{Partitioning the Data}
\begin{figure}[t]
\includegraphics[width=\columnwidth]{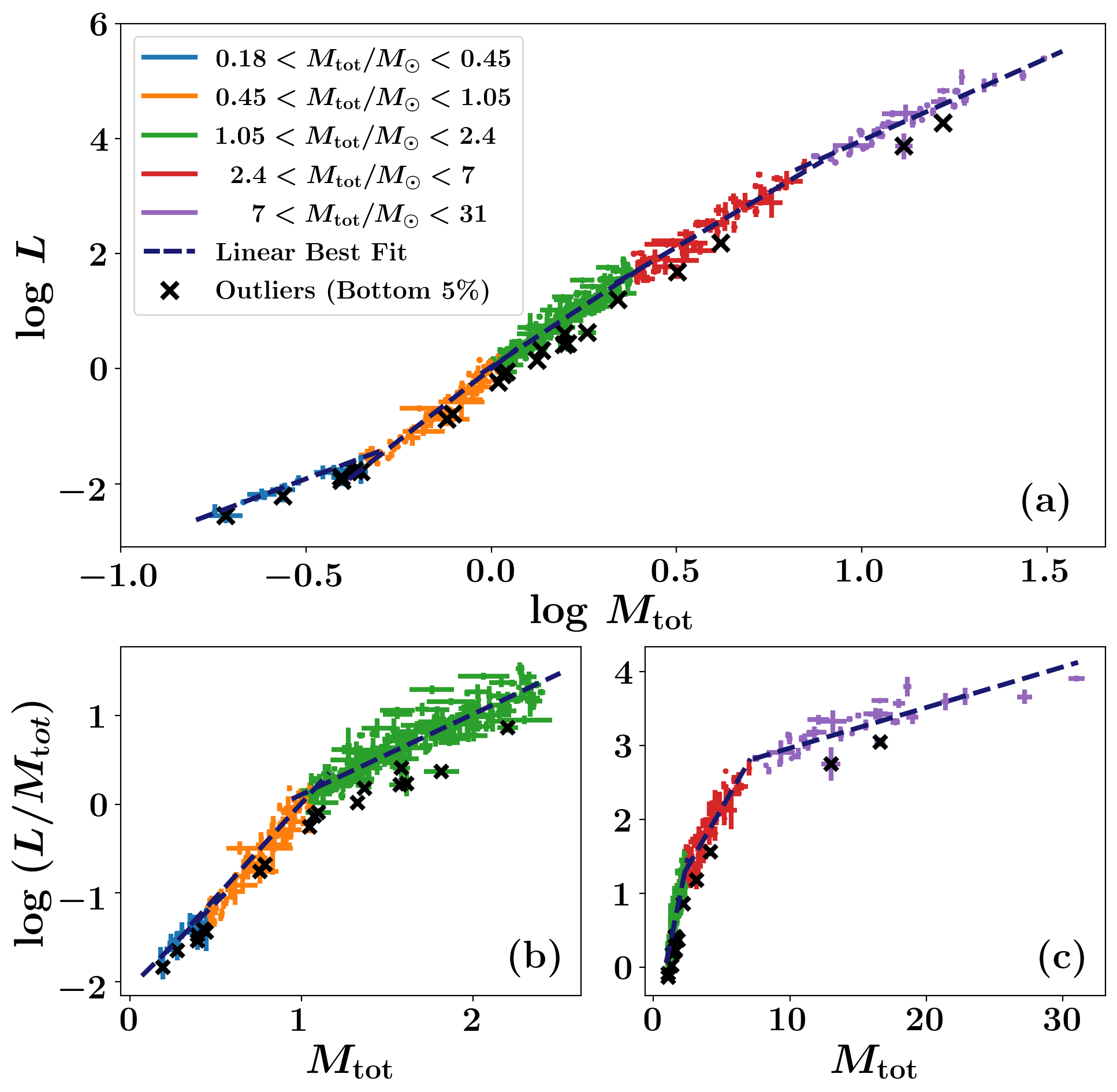}
\caption{\footnotesize Sample data and segmentation by total gravitational mass domains (reproduced from \cite{EkerData2,EkerData3}). \textbf{Panel (a):} Data is plotted in a $M-L$ diagram, in colors corresponding to partition segments. Each segment has a distinct linear best fit, and thus a distinct MLR, represented by dashed lines. Crosses mark data that are at least $2\sigma$ below the best fit, and are therefore assumed in our analysis to have null DM content.  \textbf{Panels (b), (c)}: The lower three and upper three mass ranges are each plotted on a $L/M$ axis, a measure of stellar energy production efficiency. Separate linear distributions for each domain, keeping to the panel (a) color schema, are clear. Crossed-out data points are the same ones as in panel (a). }
\label{fig:Data with segments}
\end{figure}

When placing the data on a $\log M - \log L$ plane, as shown in Fig.~\ref{fig:Data with segments}(a), 
one may observe distinct ranges of varying slope within the sample. In order to deal with this variance in a concrete manner, we adopt almost entirely the sample partition into mass regimes as suggested by Eker et al.~in~\cite{EkerData3}.

The breakpoints delineating these regimes appear to be due to changes in the energy generation mechanisms in the star, stemming either from differing dominant fission reactions in different mass ranges~\cite{MLRDerivHomol1}, or from the impact of different metallicities on the star's energy generation and evolution~\cite{SB2Basics1,SB2Basics2,PARSECmodel}. In Fig.~\ref{fig:Data with segments}(b),(c), where we reproduce the work done in~\cite{EkerData2}, we plot the same sample data on the $L/M$ axis 
(which is a measure of the stellar energy generation efficiency), showing linear distributions within distinct mass ranges with prominent breakpoints between them.

We depart from the suggested partition in one place, foregoing one of the possible breakpoints and combining two of the lower mass ranges into one. This is due to the requirements of our analysis, explored more fully throughout App.~\ref{appendix:Data Analysis Appendix}. Of the 5 remaining segments, the most heavily-populated one by some margin is the one at the~$1.05 < M_{\rm tot}/M_\odot < 2.4$ mass regime, numbering 263 stars, more than half the sample.
 
\section{Data Analysis} \label{Analysis and results}
We construct 95\% C.L.\ upper bounds by using a profile likelihood ratio (PLR) test~\cite{PLR1,PLR2}. Full details appear in App.~\ref{appendix:Data Analysis Appendix}, with the salient points summarized here.

In addition to the the effects of statistical errors in observations, each of the model parameters has a physical width due to the variance between stars, which we assume to be Gaussian. We represent this by writing our model, Eq.~\eqref{eq:DM-MLR for Pgas}, in the form,
\begin{eqnarray}
\log L + \delta \log L &=& \log (\eta + \delta \eta) + \nonumber \\ &+& 
\omega \log (Q + \delta Q) + \nonumber \\ &+& (\alpha + \delta \alpha)\log (M_{\rm tot} + \delta M_{\rm tot}) \,,
\end{eqnarray}
such that for each stellar parameter we add a Gaussian random nuisance variable, marked with $\delta$s, whose widths account for the stellar variabilities.
This allows the model parameters to obtain different values for each star in the sample. We specifically exclude $\omega$ from this treatment to avoid degeneracy issues, as explained below.

Calculating the likelihood integral for any given set of parameter values produces the 95\% C.L.\ upper bounds. The parameter space is explored via MCMC, implemented using the \textsf{emcee} Python package~\cite{emcee}.

Our MLR model, Eq.~\eqref{eq:DM-MLR for Pgas}, exhibits two distinct flat-directions, as the luminosity depends on a single combination of $\eta$, $Q$ and $\omega$.~We take the following steps to disentangle these parameters: firstly, since $\omega$ is found to be roughly constant in each of the mass regimes (see App.~\ref{appendix:Eddington's Quartic}), we confine it to a single value per regime. Secondly, seeing as even for the largest stars in our sample gas pressure 
is expected to overshadow 
radiation pressure
(again see App.~\ref{appendix:Eddington's Quartic}), we take a relatively strict lower bound prior for $\omega$, limiting its possible values to $0.8 \leq \omega \leq 1$. This disallows $Q$ from becoming arbitrarily large.
Lastly, in order to disentangle $\eta$ and $Q$, we note that the presence of DM would act to raise the luminosity. Therefore, in each mass regime we regard stars that have a $2\sigma$ deviation below the linear fit as dominated by the SM contribution, and assign these outliers with null DM content. The outliers are marked with crosses in Fig.~\ref{fig:Data with segments}.  Consequently, the $\eta$ likelihood distribution is influenced by the entire sample while the $Q$ distribution depends on a subset, removing degeneracy between them.   For more details see App.~\ref{appendix:Outlier Data}.

\begin{figure}[t]
\includegraphics[width=\columnwidth]{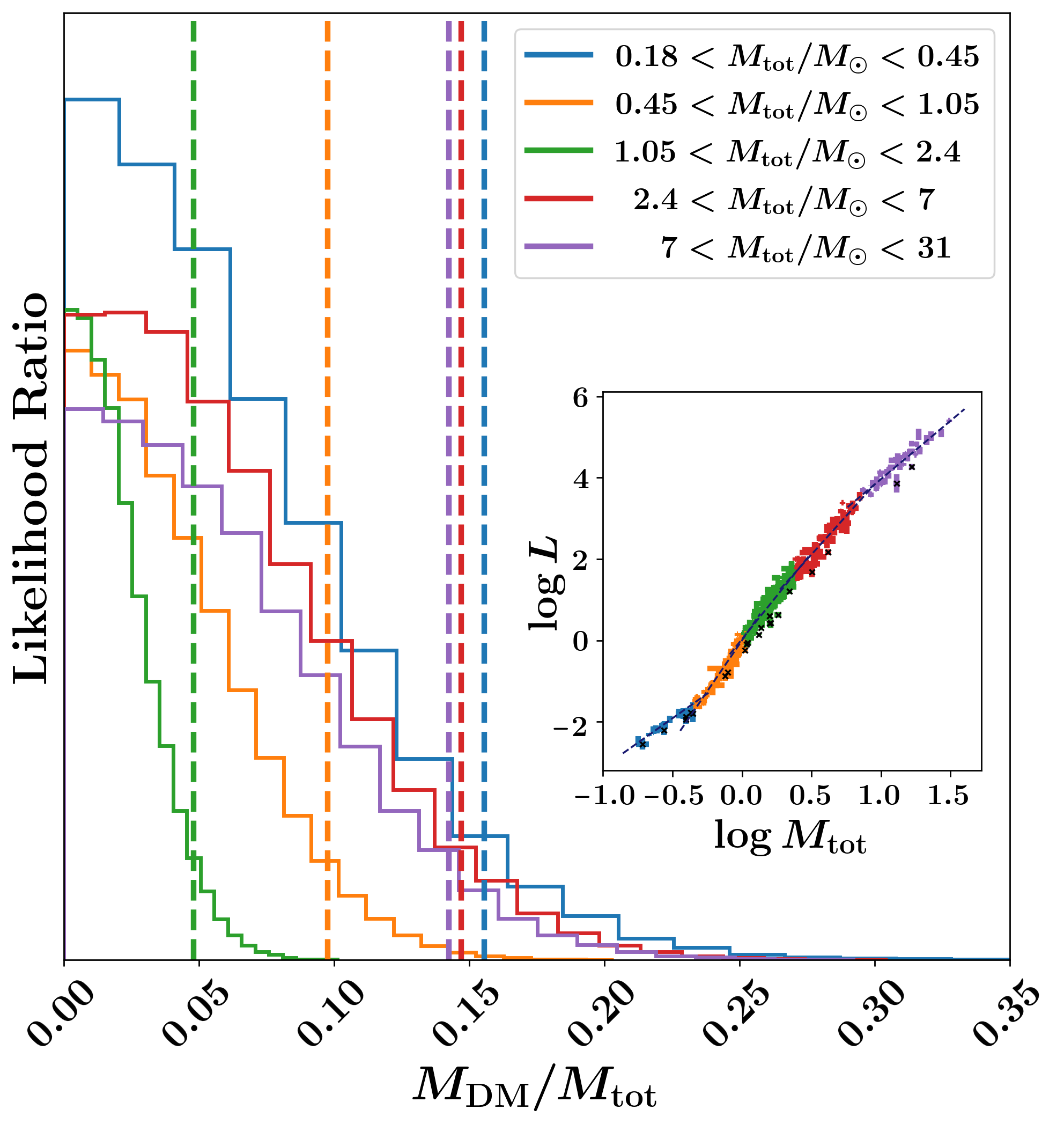}
\caption{\footnotesize Likelihood ratio histograms ({\bf solid} lines) and 95-percentile lines ({\bf dashed} vertical lines) for each of the data sample's mass ranges. Ranges are differentiated by color with the same schema as in Fig.~\ref{fig:Data with segments}, reproduced in the inset.}
\label{fig:PLR results}
\end{figure}

\section{Results}
Our MCMC analysis yields a 7-dimensional likelihood ratio map for each segment of the sample, which can be projected onto each of the individual parameters' axis after marginalizing over all the others. This results in a likelihood histogram for each parameter, and thus directly an upper bound estimation on its values. 
Taking the likelihood histograms achieved for our $Q$ parameter for each segment and translating them into $M_{\rm DM} / M_{\rm tot}$ fractions, we obtain the results of Fig.~\ref{fig:PLR results}. 
The distributions (solid lines) and 95\% C.L.~constraints (dashed vertical lines) for each mass regime are presented in the same color schema as in Fig.~\ref{fig:Data with segments}, shown again in the inset.

We observe $95\%$ C.L.\ upper bounds of at most $15\%$ DM mass across the sample, with a sub-$10\%$ upper bound in the $0.45 < M_{\rm tot}/M_\odot < 1.05$ mass regime (for which we have 82 sampled stars), and a sub-$5\%$ upper bound in the $1.05 < M_{\rm tot}/M_\odot < 2.4$ mass regime (263 sampled stars).   We comment that for the latter mass range radiative diffusion is considered to be dominant throughout the star, and thus the analytic description of the MLR is on solid grounds (see App.~\ref{appendix:Homologous Stellar Models}).
For all mass domains we find the $\omega$  distribution to be nearly constant across the allowed range (see App.~\ref{appendix:Parameter Space}).

The numerical results for the significant parameters for each of the mass domains are compiled in Table~\ref{tab:Results table}.
\begin{table}
\begin{ruledtabular}
\begin{tabular}{cccc}
Mass Range  & $N_\star$  & $\alpha $ &  ${M_{\rm DM}/M_{\rm tot}}$ Bound \\ \hline
\footnotesize ${0.18 < M_{\rm tot}/M_\odot < 0.45}$ & \footnotesize 22  & \footnotesize $2.381^{+0.112}_{-0.127}$  & \footnotesize 0.156              \\
\footnotesize ${0.45 < M_{\rm tot}/M_\odot < 1.05}$ & \footnotesize 82  & \footnotesize $5.056^{+0.098}_{-0.099}$  & \footnotesize 0.098              \\
\footnotesize ${1.05 < M_{\rm tot}/M_\odot < 2.4 \ }$  & \footnotesize 263  & \footnotesize $4.312^{+0.049}_{-0.050}$ & \footnotesize 0.048           \\
\footnotesize ${ \, 2.4 < M_{\rm tot}/M_\odot < 7 \ \ }$ & \footnotesize 77 & \footnotesize $3.835^{+0.110}_{-0.112}$ & \footnotesize 0.147           \\
\footnotesize ${ \,\ \  \ 7 < M_{\rm tot}/M_\odot < 31 \ }$ & \footnotesize 42 & \footnotesize $2.819^{+0.091}_{-0.090}$ & \footnotesize 0.143        \\
\end{tabular}
\end{ruledtabular}
\caption{\footnotesize Main results of the likelihood ratio test are summarized. For each total gravitational stellar mass domain in the sample, the  number of data points within the range ($N_\star$), the maximum likelihood power for the MLR power law ($\alpha$ in $L \propto M^\alpha)$ with $1\sigma$ limits, and the achieved $95\%$ C.L.\ upper bound for the DM mass fraction distribution, are given.}
\label{tab:Results table}
\end{table}


\section{Discussion and Outlook} \label{Discussion and Outlook}
We present a novel procedure to constrain the amount of DM within stellar interiors, utilizing a DM-motivated modification of the classic MLR and high-accuracy measurements of SB2 stars’ gravitational mass and luminosity.  We find a 95\% C.L.\ upper bound of sub-5\%  on the fractional DM mass, extracted from the main body of  data points at the $1.05<M_{\rm tot}/M_\odot < 2.4$ stellar mass range.

This result, while being noteworthy in itself, also opens some avenues for future research.  
First, analyzing a much larger stellar catalogue such as GAIA would likely yield significantly stronger constraints.  Such a study requires a dedicated analysis of accurately measured Main Sequence stars.
Second, a theoretical study that relates these limits on the parameters of dissipative DM models is highly needed.  In particular, these limits may be used to place strong bounds on the formation of other compact dark objects, such as the double disk scenario~\cite{DDMlimit}. These studies of the full dissipative DM picture are postponed to future work, to which the upper bounds we present here for the current-day total DM mass within stars may be a useful tool.

\section{Acknowledgements}
We thank Itay Bloch, Omer Katz, Tsevi Mazeh, Nadav Outmazgine and Dovi Poznanski for enlightening discussions.  This work is supported by the Israel Science Foundation (grant No. 1862/21), by the Binational Science Foundation (grant No. 2020220) and by the European Research Council (ERC) under the EU Horizon 2020 Programme (ERC-CoG-2015 - Proposal n. 682676 LDMThExp).

\bibliographystyle{apsrev4-2}
\bibliography{references.bib}

\pagebreak
\appendix
\onecolumngrid
\numberwithin{equation}{section}

\section{Derivation of the Dark Matter-Modified MLR Model}  \label{appendix:Model Derivation Appendix}
We present here the full derivation of the MLR model.
In what follows, we utilize the homologous model of stellar structure to derive an analytic description for the MLR, closely following the classic derivations presented in~\cite{MLRDerivHomol1,MLRDerivHomol2}. In Sec.~\ref{appendix:Basic equations} we present the five fundamental stellar equations used to describe evolution and structure. In Sec.~\ref{appendix:Homologous Stellar Models} we use the homologous model to arrive at the DM-modified MLR for each of the possible descriptions of the stellar interior, which vary according to its mass, dominant energy transfer mode and dominant pressure component. To complete this discussion, in Sec.~\ref{appendix:Eddington's Quartic} we study Eddington's quartic equation, used to determine the dominant pressure experienced by baryons in the star.

\subsection{Basic Equations} \label{appendix:Basic equations}
We consider a simple model of a star: spherically-symmetric, in hydrostatic and thermal equilibrium, and with negligible effects from external tidal forces or magnetic fields. Under these assumptions, stellar structure and evolution models are based on 5 equations:
hydrostatic equilibrium, mass continuity, energy transport, thermal equilibrium and the equation of state.
We re-derive these equations for a star with a non-negligible population of DM, distributed at some volume around the stellar core. For reasons that will become clear immediately, we begin with mass continuity:

\begin{enumerate}
\item {\bf Mass continuity}.
Under our assumptions, the forces acting on mass elements arise only from pressure and gravity, and as such are spherically symmetric. Therefore, for a thin shell of baryonic matter at radius $r$, width $\dd r$ and density $\rho_b$, the baryonic mass contained within it  at some time $t$ is
\begin{equation}
\dd m_b  = 4\pi r^2 \rho_b \dd r - 4\pi r^2 \rho_b v \dd t\, ,
\end{equation}
and so
\begin{equation}
\label{eq:masscont}
\pdv{r}{m_b} = \frac{1}{4\pi r^2 \rho_b} \, .  
\end{equation}

\item {\bf Hydrostatic equilibrium}.
Again for a thin shell of baryonic matter, per unit area its mass is $\rho_b\dd r$ and the gravitational force on it is $-g\rho_b\dd r$. Also per unit area, the force due to pressure differential is $\Delta P = -\pdv{P}{r}\dd r$. \\
Under hydrostatic equilibrium, these cancel out to give
\begin{equation} \label{eq:gravity}
\pdv{P}{r} = -g\rho_b \,. 
\end{equation}
Considering the added gravitational effect of DM, $g$ is 
\begin{equation} \label{eq:g_def}
g 
= \frac{G_N q(r) m_b(r)}{r^2}\,,
\end{equation}
where $q(r) \equiv {m_{\rm tot}(r)}/{m_b(r)}$. Using Eq.~\eqref{eq:masscont} we arrive at
\begin{equation}
\label{eq:HE}
\pdv{P}{m_b} = -\frac{G_N q(r) m_b(r)}{4\pi r^4}\,. 
\end{equation}

\item {\bf Energy transport}.
Energy transport in the star is ultimately a function of the temperature gradient, usually written as $\nabla = \pdv{\ln T}{\ln P}$. Using Eq.~\eqref{eq:HE}, we can write a general form for the energy transport equation,
\begin{equation}
\label{eq:Etransport}
\pdv{T}{m_b} = -\frac{G_N q(r) m_b(r)}{4\pi r^4} \pdv{T}{P} = -\frac{T}{P}\frac{G_N q(r) m_b(r)}{4\pi r^4} \nabla\,.
\end{equation}
An exact form of this equation can be written for each transport mechanism, with $\nabla_{\rm rad} = \qty(\pdv{\ln T}{\ln P})_{\rm rad}$ customarily representing transport by radiation and conduction, $\nabla_{\rm ad} = \qty(\pdv{\ln T}{\ln P})_{\rm s}$ for convective transport, and $\nabla$ itself equalling the smaller of those two.
The form of the equation for each mechanism is as follows: 

\begin{itemize}
\item {\it Radiative Diffusion}.
We assume here that DM has little effect on opacity, and that energy transport through diffusion of radiative energy is dominant over energy transport through convection.
The mean free path of photons in stars is very small compared to the stellar radius, so we can treat the photon's movement as a diffusive process. 

Let us consider some particle density $n = n(\va{x})$ moving at some average velocity $\va{v}$. Assuming isotropic motion, roughly a third of the particles would be moving in the $z$-direction, either up or down.  Consequently, the average current in that direction is obtained from the difference in up and down motion, normalized with a $1/6$ coefficient. The corresponding particle flux  over a distance $\lambda_{\rm mfp}$ at the vicinity of some height $z_0$ would thus be
\begin{align}
J_z &\simeq \frac{1}{6}\abs{\va{v}}\left[n(z_0-\lambda_{\rm mfp}) - n(z_0+\lambda_{\rm mfp})\right]
= -D \pdv{n}{z}\,,
\end{align}
with $D=\frac{1}{3} \abs{\va{v}} \lambda_{\rm mfp}$.
More generally, one has
\begin{equation}
J = - D \nabla{n} \, .
\end{equation}

To obtain the corresponding diffusive flux of radiative energy, $F$, 
we replace $n$ with the radiation energy density $u$, set $\abs{\va{v}}$ to $c$, and take the mean free path to be that of photons, $\lambda_{\rm mfp} = \frac{1}{\kappa \rho_b}$, where $\kappa$ is the opacity which has the general form
\begin{equation} \label{eq:kappa}
\kappa = \kappa_0 \rho_b^e T^f  
\end{equation}
(we neglect any opacity resulting from DM). 
Using the black-body relation $u = a T^4$ one finds
\begin{equation}
F = -\frac{4a}{3} \frac{T^3}{\kappa \rho_b} \pdv{T}{r} \,,
\end{equation}
and replacing flux with the local luminosity $l = 4\pi r^2 F$ gives
\begin{equation}
\pdv{T}{r} = -\frac{3}{16\pi a} \frac{\kappa \rho_b l}{r^2 T^3} \,.
\end{equation}
Once again using Eq.~\eqref{eq:masscont} we can write the energy transport Eq.~\eqref{eq:Etransport} with $\nabla = \nabla_{\rm rad}$ where,
\begin{gather}
 \nabla_{\rm rad} = \frac{3}{16\pi a} \frac{\kappa_0\rho_b^e T^f l P}{G_N q(r) m_b(r) T^4} \,.
\end{gather}

\item {\it Convection}.
An exact form for $\nabla_{\rm ad}$ does not exist. Some models, such as mixing-length theory~\cite{MLRDerivHomol1,StellarLifetime1}, allow for approximations of the gradient given various parameters of stellar composition and state, but in general calculations of the convective gradient are not analytic. 
Fortunately, a reasonable approximation exists with the much simplified assumption of a polytropic stellar model~\cite{MLRDerivHomol2}, with index $\gamma$, for which  
the gradient is simply constant, $\nabla_{\rm ad} = \frac{\gamma - 1}{\gamma}$. In the case of ideal gas, for instance, this gives $\nabla_{\rm ad} = 0.4$.
We therefore take for Eq.\eqref{eq:Etransport} $\nabla=\nabla_{\rm ad} = $ const. 
\end{itemize}

\item {\bf Thermal equilibrium}.
We neglect here the energy released from DM through mechanisms such as annihilation, assuming that at the Main Sequence stage we are past the period in the star's evolution at which these mechanisms, if contributory, had a substantial effect. We define $\epsilon$ as the energy released per unit baryonic mass per second, which has the general form 
\begin{equation}
\label{eq:epsilon}
\epsilon=\epsilon_0 \rho_b^g T^h\,. 
\end{equation}
$l(r)$, the local luminosity, is the net energy per second passing through a sphere with radius $r$.
For a thin shell of baryonic matter at radius $r$, $\dd l = 4\pi r^2 \rho_b \epsilon \dd r = \epsilon \dd m_b$, and therefore
\begin{gather}
 \pdv{l}{m_b} = \epsilon \,.
\end{gather}

\item {\bf Equation of state}.
We refer here to the pressure experienced by baryons in the star, and so this is simply the usual ideal gas pressure with the addition of radiation pressure. The relative prominence of each of those will affect the form of our model.
\begin{equation} \label{eq:EoS}
P = P_{\rm gas}+P_{\rm rad} = \frac{k_B}{m_p\mu}\rho_b T + \frac{1}{3}aT^4 \,.
\end{equation}

\end{enumerate}

With the above derivations  we arrive at Eqs.~\eqref{eq:basic_HE}-\eqref{eq:basic_EoS}.  To make further progress with deriving constraints on the presence of DM within stars, one must arrive at a general form of the MLR. Below we employ homology to arrive at such a form  to be used in the analysis.

\subsection{Homologous Stellar Models} \label{appendix:Homologous Stellar Models}
A homologous (or "self-similar") stellar model enables us to reduce the differential stellar structure equations into algebraic ones, and receive relations between the stellar variables through dimensional analysis. 
The central concept of these homology models is that they describe stars with internal structures that are in some way self-similar. Specifically, closely following the notations of~\cite{MLRDerivHomol2}, consider a pair of stars whose structures are described by unprimed and primed variables. In this pair of stars, we can identify homologous baryon mass shells as having the same relative mass coordinate in each star, 
\begin{equation}
\xi = \frac{m_b}{M_b} = \frac{m'_b}{M'_b} \,.
\end{equation}
We then say that the stars are homologous if the relative radii of the homologous mass shells are also equal, that is, for all $\xi$,
\begin{equation}
\frac{r(\xi)}{R} = \frac{r'(\xi)}{R'} \quad \Rightarrow \quad \frac{r(\xi)}{r'(\xi)} = \frac{R}{R'} = \text{const.}
\end{equation}
As a consequence,  there is some dimensionless function $f(\xi)$ such that $r = f(\xi) R$ for all homologous stars. 

Since a pair of homologous stars must both obey the stellar structure equations, there should be definite relationships between all their parameters, that similarly scale in the same way for all stars for some function of $\xi$.
We find these relationships through dimensional analysis,
\begin{subequations}
\begin{align}
P &= f_1(\xi) P_\star \label{eq:decomp subeq1} \,,\\
r &= f_2(\xi) R_\star \label{eq:decomp subeq2} \,,\\
\rho_b &= f_3(\xi) \rho_{b\star} \,,\\
T &= f_4(\xi) T_\star \,,\\
l &= f_5(\xi) L_\star \,,\\
q &= f_6(\xi) Q_\star \,, \label{eq:decomp subeq6}
\end{align}
\end{subequations}
such that the $f$s are dimensionless functions of $0 \leq \xi \leq 1$. This separation has another implicit assumption - that each star is chemically homogeneous, i.e., has a constant mean molecular mass $\mu$. This also implies that $\kappa, \epsilon$ have the same functional shape throughout the star. We can now use these to get algebraic relations between the stellar parameters. Taking for example Eq.(\ref{eq:HE}) and substituting Eqs. (\ref{eq:decomp subeq1}),(\ref{eq:decomp subeq2}) and (\ref{eq:decomp subeq6}) we get
\begin{equation}
\frac{P_\star}{M_b} \pdv{f_1}{\xi} = -\frac{f_6 G_N Q_\star \xi M_b}{f_2^4 R_\star^4}\,,
\end{equation}
from which it follows, without loss of generality, that,
\begin{equation}
  \begin{split}
    \pdv{f_1}{\xi} = \frac{f_6\xi}{f_2^4} 
  \end{split}
 \quad \quad , \quad \quad
    \begin{split}
    P_\star = \frac{G_N Q_\star M_b^2}{R_\star^4} \,. 
  \end{split}
\end{equation}
We repeat this procedure for each of Eqs.~\eqref{eq:basic_MCont},\eqref{eq:basic_TE} and \eqref{eq:basic_EoS}, to arrive at a scaling relation for either gas or radiation  pressure dominance in the stellar interior, 
\begin{equation}
  \begin{split}
    \pdv{f_1}{\xi} &= \frac{f_6\xi}{f_2^4} \\
    \pdv{f_2}{\xi} &= \frac{1}{f_2^2 f_3 } \\
    \pdv{f_5}{\xi} &= f_3^g f_4^h \\
    f_1  = &\left\{ 
    \begin{array}{llll}
    f_3 f_4 &&& 
    \\
    f_4^4 &&& 
    \end{array}\right.
  \end{split}
 \quad \quad \quad
    \begin{split}
    P_\star &= \frac{G_N Q_\star M_b^2}{R_\star^4}\, \hspace{10.2em} , \\
    \rho_\star &= \frac{M_b}{R_\star^3}\, \hspace{13.15em} , \\
    L_\star &= \epsilon_0 \frac{M_b^{g+1}}{R_\star^{3g}} T_\star^h \, \hspace{9.85em} ,\\
    T_\star &= \left\{ 
    \begin{array}{llll}
    \frac{\mu P_\star R_\star^3}{M_b} &&& \qquad\textrm{   $P_{\rm gas}$ dominant} \\
    P_\star^{\frac{1}{4}} &&&\qquad \textrm{  $P_{\rm rad}$ dominant}
    \end{array}\right.
    \,.
  \end{split}
  \label{eq:scaling relations}
\end{equation} 
Here we used the first and second terms of Eq.~\eqref{eq:EoS} for gas and radiation pressure respectively.  

We now address Eq.~\eqref{eq:basic_ET} in a similar fashion.  Our discussion in the following subsection distinguishes between radiative diffusion, in which both gas or radiation pressure can dominate, and convection, for which gas pressure necessarily dominates. 

\subsubsection{Energy transport models}
For {\bf radiative diffusion} the appropriate form of the equation leads to
\begin{equation}
  \begin{split}
    \pdv{f_4}{\xi} &= -\frac{f_3^e f_4^{f-3}f_5}{f_2^4} 
  \end{split}
 \quad \quad , \quad \quad
    \begin{split}
    L_\star &= \frac{T_\star^{4-f}R_\star^{4+3e}}{\kappa_0 M_b^{1+e}} \,.
  \end{split}
\end{equation} 
From here, the treatment diverges according to the dominant pressure mode:
\begin{itemize}
\item {\bf $P_{\rm gas}$ dominance}.
We perform some simple algebra, then return from the starred versions of variables to the physical ones, and assume the DM resides within the boundaries of the star (so that we can take for simplicity $f_6(\xi=1) = 1$ and $Q_*=Q$). This results in
\begin{gather}
L = f_5(1) \frac{(\kappa_0 \epsilon_0)^{\frac{3e+f}{3e+f+3g+h}}}{\kappa_0}
		\mu^{\frac{4(3g+h)+3(eh-fg)}{3e+f+3g+h}} G_N^{\frac{4(3g+h)+3(eh-fg)}{3e+f+3g+h}} 
		Q^{1+\frac{e(h-6)-f(2+g)}{3e+f+3g+h}} 
		M_{\rm tot}^{3+\frac{2e(h-3)-f(2+2g)}{3e+f+3g+h}} \,,
\end{gather}
where the parameters $e$, $f$, $g$, and $h$ are defined in Eqs.~\eqref{eq:kappa},~\eqref{eq:epsilon}. More succinctly,
\begin{equation}
L = \tilde{\eta}\,G_N^{\alpha+\varphi} Q^\varphi M_{\rm tot}^\alpha \,,
\end{equation}
where $\tilde{\eta}$ encompasses all of the SM physics factors excluding $G_N$ in the coefficient.
This is the general result for this stellar mode. Specifying electron scattering dominated opacity, which behaves as $\kappa = \kappa_{0,\rm es}$, i.e.~$e=f=0$, gives the familiar zeroth-order approximation for the MLR with the added $Q$ factor,
\begin{equation}
L \propto Q M_{\rm tot}^3 \,.
\end{equation}

\item {\bf $P_{\rm rad}$ dominance}.
Radiation pressure dominates for a temperature high enough such that $\frac{1}{3}aT^4 \gg\frac{k_B}{m_p\mu}\rho_b T $. We can circumvent some of the work here by noting that at such high temperatures the star is likely to be completely ionized, and so electron scattering dominates opacity. This again means $e=f=0$, which will make our calculation here easier. Using Eq.~\eqref{eq:HE},
\begin{equation}
T_\star^4 = \frac{G_N Q_\star {M_b}^2}{R_\star^4} \,,
\end{equation}  
therefore
\begin{equation}
L_\star = \frac{T_\star^4}{\kappa_{0,\rm es}} \frac{R_\star^4}{M_b} \sim \frac{G_N Q_\star {M_b}^2}{R_\star^4}\frac{R_\star^4}{M_b} = G_N Q_\star {M_b} = G_N M_{\rm tot} \,,
\end{equation}
and so
\begin{equation}
L = \tilde{\eta}\,G_N M_{\rm tot} \,.
\end{equation}
We lose the explicit dependence on $Q$ and now only have dependence on the total gravitational mass, same as in the classic case for radiation pressure dominance.
\end{itemize}

The {\bf convection} case requires a slightly more involved treatment.  Indeed, a clear analytic description from first principles of the convective stellar interiors, needed for modeling the MLR, Eq.~\eqref{eq:DM-MLR for Pgas}, does not exist.  Instead, the simple assumption of $\nabla = \nabla_{\rm ad} = \text{const.}$, relevant for convection~\cite{MLRDerivHomol1,MLRDerivHomol2},  allows one for an analytical approximation which we derive below.  We stress, however, that the general form of Eq.~\eqref{eq:DM-MLR for Pgas}, agrees both with this simplified treatment below as well as the more involved numerical studies (see~\cite{PARSECmodel} and its use in~\cite{EkerData3}). 

For a constant $\nabla$ a complication arises since Eq.~\eqref{eq:basic_ET} is degenerate with Eq.~\eqref{eq:basic_HE}. To make progress, we will instead utilize the familiar black-body relation
\begin{equation}
\label{eq:black-body}
L = 4\pi R^2\sigma_{SB} T_{\rm eff}^4 \,,
\end{equation}
and rewrite $R$ and $T_{\rm eff}$ as functions of $M_b$ to arrive at the MLR. To do so, we use three equivalent expressions for the pressure at the edge of the star, $P_R \equiv P(r=R)$, where Eq.~\eqref{eq:black-body} holds.  These expressions will result with the needed relations. 

A first expression for $P_R$ can be obtained from the polytropic stellar model with index $n=3$~\cite{MLRDerivHomol1,MLRDerivHomol2}, 
\begin{equation} \label{eq:Ppoly}
    P = K \rho_b^{1+\frac{1}{n}}\,, 
\end{equation}
where
\begin{equation}
K^n = c_1(G_N Q)^n M_b^{n-1} R^{3-n}\,,
\end{equation}
with $c_1$ a constant.
We thus have, 
\begin{gather}
P_R = c_1 \frac{G_N Q M_b^2}{R^4} \,.
\end{gather}
A second expression stems from Eqs.~\eqref{eq:gravity} and~\eqref{eq:g_def}:
\begin{equation}
P_R = \frac{G_N Q M_b}{R^2} \int_R^\infty \rho_b\,dr  \,,
\end{equation}
and using the fact that (by definition) at $r=R$ the optic depth, $\tau$, is equal to 1,
\begin{equation}
1 = \tau_R = \int_R^\infty \kappa \rho_b\,dr = \kappa_R \int_R^\infty \rho_b\,dr = \kappa_0 \rho_{b_R}^e T_{\rm eff}^f \int_R^\infty \rho_b\,dr \,,
\end{equation}
we get, again for some constant $c_2$,
\begin{equation}
P_R 
    = c_2 \frac{G_N Q M_b^{1-e}}{R^{2-3e}} T_{\rm eff}^{-f} \,.
\end{equation}
Finally, for convection, gas pressure necessarily dominates and thus the appropriate equation of state  at the surface is
\begin{equation}
P_R = \frac{k_B}{\mu m_p} \rho_b T_{\rm eff} = c_3 \frac{M_b}{R^3} T_{\rm eff} \,.
\end{equation} 

Equating each two of the above three expressions for $P_R$ we obtain (up to constant coefficients):
\begin{subequations}
\begin{gather}
T_{\rm eff}^{2+f} \propto G_N Q\, \frac{R^{3e+1}}{M_b^e} \,,\\
R^{2+3e+f} \propto (G_N Q)^f\, M_b^{1+e+f} \,, \\ 
T_{\rm eff}^{2+3e+f} \propto (G_N Q)^{2+3e}\, M_b^{1+2e}\,.
\end{gather}
\end{subequations}
Substituting these in Eq.~\eqref{eq:black-body} yields
\begin{equation}
L = \tilde{\eta}\, G_N^{4-\frac{2f}{2+3e+f}} Q^{1-\frac{e+f}{2+3e+f}} M_{\rm tot}^{3+\frac{e-f}{2+3e+f}} \,.
\end{equation}
This result has the same structure as in the other stellar modes, and indeed for opacity dominated by electron scattering we once again find
\begin{equation}
L \propto Q M_{\rm tot}^3\,.
\end{equation}
However, for Kramers opacity, which is expected to be at least prominent in convection-dominated lower-mass stars, we get
\begin{equation}
L \propto Q^\frac{8}{3} M_{\rm tot}^6 \,. 
\end{equation}
This higher power for the mass is in contention with the clear decrease in slope observed at the lower ends of the MLR diagram. Thus we see that the approximations we rely on for the analytic derivations are less reliable for the convection-dominated mode. A more accurate model requires a more in-depth approach for the effects of DM on convective stellar interiors, which is outside the scope of this work.

\subsubsection{A cohesive form for the MLR}
In all modes we obtain an MLR of the general form
\begin{equation}
L = \tilde{\eta}\,G_N^{\alpha+\varphi} Q^\varphi M_{\rm tot}^\alpha \,.
\end{equation}
However, while the power of $Q$ is greater or equal to one in the diffusive, gas pressure dominated scenario as well as in  convection-dominated stars, in radiation pressure dominated ones it is, in fact, equal to 0. 

In order to work with a model incorporating all of these features, we write our DM-modified MLR in the following general form:
\begin{equation}
L = \eta\,\qty(Q^\omega)^\phi M_{\rm tot}^\alpha  \,,
\label{eq:full DM-MLR model}
\end{equation}
with $\phi \geq 1$ and $0 \leq \omega \leq 1$. Furthermore, in order to avoid degeneracy and ensure a conservative estimate for the upper bound on $Q$, we set $\phi = 1$, which results with the MLR model of Eq.~\eqref{eq:DM-MLR for Pgas}.
$\omega$ is determined for each star by the relative impacts of the gas and radiation pressures; to find this we must consider the Eddington quartic equation, to be discussed in the following subsection. 
Before that, however, a note on the model validity is in order.

\subsubsection{Model validity}
The lack of $Q$-dependence in the $P_{\rm rad}$-dominated case serves as a natural cap for a seeming oddity in the model for non-zero $\omega$ values in Eq.~\eqref{eq:full DM-MLR model}, in which arbitrarily raising the value of $Q$ (i.e.~arbitrarily raising the relative amount of DM mass to baryon mass) would cause luminosity to rise, which would seem unreasonable. However, the dependence of $T$ on $Q$ is of the form (see Eq.~\eqref{eq:basic_ET})
\begin{equation}
T \sim \qty(\frac{G_N Q M_b^2}{R^4})^\nu \,,
\end{equation}
with $\nu$ being some positive power resulting from the equation of state. This means that increasing $Q$ would  result in a more pronounced presence of radiation pressure, which ultimately means a diminished dependence of $L$ on $Q$. That is, $L$ would not tend to infinity even if $Q$ does. 

This solution prevents the luminosity from arbitrarily rising, but is otherwise not a very strong limit on the influence of a high $Q$ value, as radiation pressure becomes dominant only when the star's mass is several orders of magnitude larger than $M_\odot$. We can find a stronger limit in the stellar lifetime: the largest timescale for the life of a star on the Main Sequence is the nuclear timescale~\cite{MLRDerivHomol1},
\begin{equation}
\tau_{\rm nuc} = \frac{\varepsilon M}{L} \,,
\end{equation}
where $\varepsilon$ is some efficiency factor. Using our notation, and noting that energy is specifically generated by the baryon mass, we find (for electron scattering dominated opacity; other cases result with an even more severe effect),
\begin{equation}
\tau_{\rm nuc} \sim \frac{\varepsilon M_b}{Q M_{\rm tot}^3} = \frac{\varepsilon M_b}{Q^4 M_b^3} \,.
\end{equation}
That is, given a set amount of baryonic matter in the star, an arbitrary increase in $Q$ would result in a quartic decrease in its lifetime on the Main Sequence at the least, and an even more precipitous drop for other opacity modes. More concretely, for $Q=2$, fitting an equal amount of baryon matter and DM in the star, its lifetime would decrease by a factor between 1 and 2 orders of magnitude while  for, e.g.~$Q=6$, 
the lifetime would decrease between 3 and 6 orders of magnitude. 
The significant discrepancy this would present with current astrophysical pictures of stars' ages and evolution (see e.g.~\cite{StellarLifetime1,StellarLifetime2}) means that $Q$ is very unlikely to be more than slightly larger than 1.

\subsection{Eddington's Quartic} \label{appendix:Eddington's Quartic}
As we've seen above, $\omega$ of Eqs.~\eqref{eq:DM-MLR for Pgas} and~\eqref{eq:full DM-MLR model} is determined according to the stellar pressure composition.  Here we'd like to relate this composition to the stellar mass and DM fraction.
To this end we use Eddington's quartic equation, originally resulting from Eddington's standard stellar model, which relays the connection between a star's mass and the relative dominance of gas pressure versus radiation pressure. We re-derive it here for the DM-modified case, following the classical derivation as it appears in \cite{MLRDerivHomol2, EddingQuartic}, the modification being limited to a distinction between the total and baryonic masses, and the substitution $G_N \rightarrow G_N Q$.

\begin{figure}[t!]
\includegraphics[width=18em]{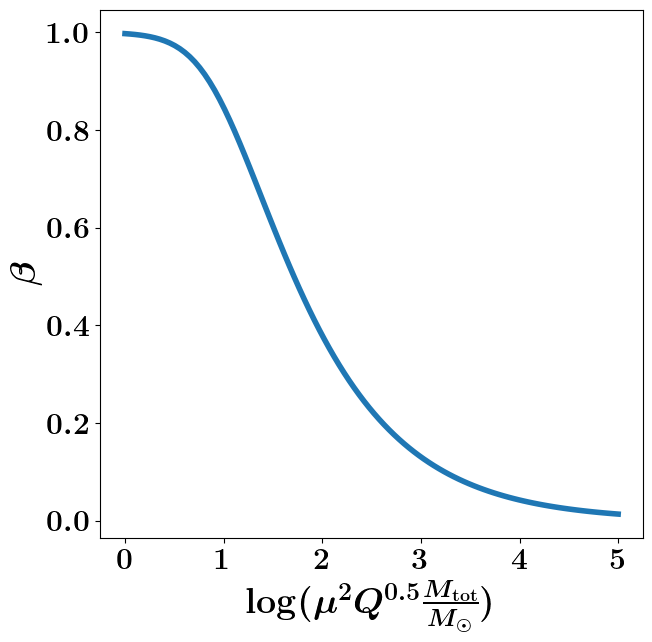}
\centering
\caption{\footnotesize Solution of Eddington's quartic equation relating the gas-to-total pressure ratio, $\beta$, to the stellar mass and dark matter fraction.   One finds that for $0< Q-1 \ll 1$, $P_{\rm gas}$ dominates ($\beta \sim 1$) for $M_{\rm tot}/M_\odot \lesssim 20$, which holds for the majority of stars studied in this paper.} 
\label{fig:Eddington Quartic}
\end{figure}

We define $\beta$ as the ratio of gas pressure to the total pressure experienced by baryons in the star.  One then assumes that $\beta$ is constant throughout the star, which implies
\begin{equation}
\label{eq:Peq}
\frac{aT^4}{3(1-\beta)} = \frac{P_{\rm rad}}{1-\beta} = P = \frac{P_{\rm gas}}{\beta} = \frac{k_B}{m_p\mu\beta}\rho_b T \,,
\end{equation}
and therefore
\begin{equation}
T = \qty[\frac{3k_B}{a m_p \mu}\frac{1-\beta}{\beta}]^\frac{1}{3}\rho_b^\frac{1}{3} \,.
\end{equation}
Substituting this back in Eq.~\eqref{eq:Peq} gives us the polytropic equation of state, Eq.~\eqref{eq:Ppoly}, with index $n=3$, and a coefficient term
\begin{equation}
K = \qty[\frac{3k_B^4}{a m_p^4 \mu^4}\frac{1-\beta}{\beta^4}]^\frac{1}{3} \,.
\label{eq:eddington_K}
\end{equation}

Using Eq~\eqref{eq:Ppoly} with the hydrostatic equation,  Eq.~\eqref{eq:HE}, one arrives at a modified identity for polytropic stars,
\begin{equation}
\qty(\frac{G_N Q M_b}{M_n})^{n-1} \qty(\frac{R}{R_n})^{3-n} = \frac{[(n+1)K]^n}{4\pi G_N Q} \,,
\end{equation}
with $M_n, R_n$ being constants dependent on the polytropic index. Since $n=3$ this results in
\begin{equation}
M_b^2 = (4\pi M_3)^2 \qty(\frac{K}{\pi G_N Q})^3 \,.
\end{equation}
Reintroducing Eq.~\eqref{eq:eddington_K} gives us
\begin{gather}
\mu^4 Q M_{\rm tot}^2 = \frac{3k_B^4 (4\pi M_3)^2}{a m_p(\pi G_N)^3}\frac{1-\beta}{\beta^4} \,,
\end{gather}
and thus
\begin{equation}
\mu^4 Q \qty(\frac{M_{\rm tot}}{M_\odot})^2 = 0.003\frac{1-\beta}{\beta^4} \,.
\end{equation}
This is the DM-modified Eddington's quartic equation, whose solution is shown in Fig.~\ref{fig:Eddington Quartic}.

To give some order-of-magnitude guidance, we assume the solar value for the mean molecular weight, $\mu=0.6$, and that $Q$ is of order 1. For these values we have clear $P_{\rm gas}$-dominance ($\beta \geq 0.9$) up until $M_{\rm tot}/M_\odot \sim 20$; equivalent contributions from each pressure term ($\beta = 0.5$) is found for $M_{\rm tot}/M_\odot \sim 150$; and clear $P_{\rm rad}$-dominance ($\beta \leq 0.1$) is obtained starting from $M_{\rm tot}/M_\odot \sim 5000$. 
Since the largest stellar mass in our sample is $M_{\rm tot}/M_\odot =31$ (corresponding to $\beta = 0.83$, indicating a strong prominence of the $P_{\rm gas}$ term), we take the gas-pressure mode as the dominant one for our analysis, thus limiting the range of the above mentioned $\omega$ factor close to unity in our Bayesian priors, $0.8<\omega<1$.

\section{Statistical Analysis} \label{appendix:Data Analysis Appendix}
\subsection{Likelihood Calculation} \label{appendix:Likelihood Calculation}
We constain the presence of DM within stars  using a profile likelihood ratio (PLR) test (see e.g.~\cite{PLR1,PLR2}). 
Our likelihood function takes into account not only statistical uncertainties but also the physical variance in stellar parameters, expected due to the natural variation of stellar composition and structure.  
Thus for each of the parameters $\eta$, $Q$, and $\alpha$ of Eq.~\eqref{eq:DM-MLR for Pgas}, we add a Gaussian random nuisance variable, $\delta \eta$, $\delta Q$ and $\delta \alpha$ respectively, with widths accounting for the stellar variabilities, which we fit for using the PLR test. We further add two nuisance parameters, $\delta M$ and $\delta\log L$ (both Gaussian distributed), that encapsulate the physical variability and systematic uncertainty in extracting the mass and luminosity.  However, the width of these variables are not left free and are taken directly from the error bars of the reported data, which take into account systematic uncertainties (see~\cite{EkerBothMeas1} for example).   Concretely,
\begin{equation}
\log L + \delta \log L = \log (\eta + \delta \eta) + \omega \log (Q + \delta Q) + (\alpha + \delta \alpha)\log (M_{\rm tot} + \delta M_{\rm tot}) \,.
\end{equation}
We stress that, with the above, all of the $\delta$ variables are allowed to vary between stars.  The original parameters ($\eta$, $Q$, $\alpha$, $M$ and $\log L$) now represent (with a mild abuse of notation) the means of the parameter distributions, since we assume a locally Gaussian noise. Given infinite data with complete accuracy, the variance parameters would be equal to the physical variance.  However, since we do not have infinitely accurate data, they also encompass the statistical errors of sampling. 

The $\omega$ parameter is excluded from this treatment due to degeneracy issues. Its physical significance as representing the degree to which $P_{\rm rad}$ is prominent is also one that is likely not significantly different between stars within the same mass domain, and therefore it can be safely fitted for each domain as a whole, rather than for each star. 

Using conditional probabilities, we can therefore write the likelihood in the following generic form:
\begin{equation}
\mathcal{L}(\theta, \sigma_\theta\,|\,\vb{d}) = \int d\delta\theta\ P(\delta\theta) P(\vb{d}\,|\,\theta, \delta\theta)
\end{equation}
where $\theta$, $\delta\theta$ are vectors of parameter means and deviations, $P(\delta\theta)$ is the probability of getting a specific deviation from the parameter distribution mean (in this case, simply a Gaussian centred around 0), and $\int d \delta\theta$ runs over all of the relevant deviations. The conditional probability is then reduced to a product of a Dirac $\delta$ function for each of the data points,
\begin{equation}
P(\vb{d}\,|\,\theta, \delta\theta) = \prod_{i=1}^{n} \delta\Big(\vb{d}_i - \theta_i - \delta\theta_i\Big) \,.
\end{equation}
A bit more explicitly, the form of our likelihood function (after making the fairly reasonable assumption that the variances for the parameters are small relative to their mean values, allowing us to make convenient approximations for the $\log$s) is as follows:
\begin{gather}
\mathcal{L}(\alpha, \sigma_\alpha, \eta, \sigma_\eta, Q, \sigma_{Q}, \omega \, |\,{\{M_{\rm tot_i}, \sigma_{M_{\rm tot_i}}, \log L_i, \sigma_{\log L_i}\}}_{i=1}^n)= \nonumber \\
= \prod_{i=1}^n \int d\delta M_{\rm tot_i}\,d\delta \alpha_i\,d\delta \eta_i\,d\delta Q_i\,d\delta \log L_i\ 
G(\delta M_{\rm tot_i} | \sigma_{M_{\rm tot_i}})\,G(\delta \alpha_i | \sigma_\alpha)\,
G(\delta \eta_i | \sigma_\eta)G(\delta Q_i | \sigma_{Q})\,G(\delta \log L_i | \sigma_{\log L_i})  \\
\delta\left(\delta \log L_i - \left[\log \eta + \frac{\delta\eta_i}{\eta} + \omega\, \left(\log Q + \frac{\delta Q_i}{Q}\right) + \left(\alpha + \delta \alpha_i\right)\left(\log M_{\rm tot_i} +  \frac{\delta M_{\rm tot_i}}{M_{\rm tot_i}}\right) - \log L_i\right]\right) \nonumber
\end{gather}
with $G(\delta X | \sigma_X) = \frac{1}{\sqrt{2\pi}\sigma_X}\,e^{-\frac{(\delta X)^2}{2\sigma_X^2}}$ for each parameter $X$. We note that the $\delta(\cdot)$ in the last line represents a Dirac $\delta$ function.  We can calculate most of these integrals directly, which gives us (for compactness, we write this for one term in the product, leaving the $i$ index outside):
\begin{align}
\mathcal{L}_i
&= \int d\delta M_{\rm tot}\,d\delta \alpha\,d\delta \eta\,d\delta Q \nonumber \\
& \frac{\exp \qty(-\left[\frac{\delta M_{\rm tot}}{2{\sigma_{M_{\rm tot}}}^2} + \frac{\delta Q}{2{\sigma_{Q}}^2} + \frac{\delta \alpha}{2{\sigma_\alpha}^2} +\frac{\delta \eta}{2{\sigma_\eta}^2} + 
\frac{\left[\left(\alpha + \delta\alpha \right) \
\left(\log M_{\rm tot} + \frac{\delta M_{\rm tot}}{M_{\rm tot}} \right) + \omega \left(\log Q + \frac{\delta Q}{Q} \right) + \log \eta + \frac{\delta\eta}{\eta}- \log L\right]^2}{2{\sigma_{\log L}}^2}\right])}
{4\sqrt{2}\,\pi^{5/2}\,\sigma_{\log L} \sigma_{M_{\rm tot}} \sigma_{Q} \sigma_\alpha \sigma_\eta} =
\nonumber \\
\nonumber \\
&= \int d\delta M_{\rm tot}\, \frac{\eta\, Q \exp \left(-\frac{1}{2} \left[\frac{{\delta M_{\rm tot}}^2}{{\sigma_{M_{\rm tot}}}^2} + \frac{\eta^2 Q^2 \left[\alpha \left(\log M_{\rm tot} + \frac{\delta M_{\rm tot}}{M_{\rm tot}} \right) + \log \eta + \omega \log Q - \log L \right]^2}{\eta^2 \sigma_Q^2 \omega^2 + \sigma_\eta^2 Q^2 + \eta^2 Q^2 \left[\sigma_{\log L}^2 + \sigma_\alpha^2 \left(\log M_{\rm tot} + \frac{\delta M_{\rm tot}}{M_{\rm tot}} \right)^2 \right]} \right] \right)}
{2\pi \sigma_{M_{\rm tot}} \sqrt{\eta^2 \sigma_Q^2 \omega^2 + \sigma_\eta^2 Q^2 + \eta^2 Q^2 \left[\sigma_{\log L}^2 + \sigma_\alpha^2 \left(\log M_{\rm tot} + \frac{\delta M_{\rm tot}}{M_{\rm tot}} \right)^2 \right]}}\,.
\end{align}
This final form for the likelihood integral can be solved numerically in Python, using standard \verb|NumPy| functions.
Setting the integration bounds between $\pm 5$ times the maximum recorded error and calculating on a 101 steps grid gave results with negligible difference to those found in \verb|Mathematica|, and there was no discernible difference to the results when larger integration bounds or a more finely-grained grid were used.

This calculation, performed for each set of proposed values for the model parameters, produces a likelihood ratio test, and using its result gives us the sought after 95\% C.L.\ upper bound for each of the parameters’ distributions. The parameter of interest here is of course $Q$, since the bound on its distribution directly gives the bound on the fractional DM content distribution for the stars in the sample.

Receiving useful results from the calculation required resolving the degeneracy between some of the model parameters, specifically $\eta$ and $Q$. The differing physical significance for each of them - one relating the SM contributions to the MLR coefficient, the other the DM contributions - does not translate into any different treatment a statistical test would give them, and so one parameter could be arbitrarily large and the other arbitrarily small while the test produces the same result. In order to create a clearer distinction between the parameters' distributions, we give a specific significance to outlying data.

\subsection{Outlier Data} \label{appendix:Outlier Data}
We introduce an additional assumption into the model that reflects its underlying physical interpretation and would create a distinction between the parameters. This is done by assigning some of the stars in our sample with null DM content.
We have constructed our model such that each star may have its own $\eta$- and $Q$-values, reflecting its composition in both the visible and dark sectors, and such that any DM content results in a higher expected luminosity value (see Eq.~\eqref{eq:DM-MLR for Pgas}). Stars with significantly low luminosity relative to what we may expect from a locally estimated MLR may therefore be assumed to have subdominant DM content.   While it is possible that a star with low luminosity would have a sizable $Q$-value, this would imply an anomalously low $\eta$-value which is inconsistent with its observed distribution among other stars in the same mass domain.


We therefore make the following arbitration: In each mass regime, those stars that have a 2$\sigma$ deviation below the mass-luminosity best-fit line are assumed to be completely dominated by the SM contribution, and are assigned to have null DM content. As this changes the fit to which we compare in the first place, we take an iterative approach to the procedure: finding a max-likelihood linear fit for the MLR; marking the samples within the lowest 5 percentiles as outliers and excluding them; re-fitting with the remaining samples and repeating until no more data points have a $2\sigma$ deviation below their expected value. For our sample, this process terminated after 2 or 3 iterations for each mass regime, excluding overall 4.5\% of the data as outliers with null DM content (22 out of 486 stars). 

As a point of note, our initial partition of the data into mass regimes had separated the $0.45 < M_{\rm tot}/M_\odot < 0.72$ and $0.72 < M_{\rm tot}/M_\odot < 1.05$ domains as suggested in \cite{EkerData3}. However, under this partition the former segment did not contain any $\geq 2\sigma$ outlying data. Noting that the two regions presented nearly identical slopes in the $\log M - \log L$ diagram, we removed the break between them, allowing for a consistent treatment across all mass regimes of our data, while not significantly impacting our results. This is the cause for our partitioning of the sample into 5 domains instead of the 6 proposed in \cite{EkerData3}.

This outlier data exclusion results in a different treatment of the $\eta$-distribution  and the $Q$-distribution parameters: for specific data points, $Q$ and its width are forced to vanish, while $\eta$ benefits from the entire data set, removing the flat direction.


\subsection{Exploring the Parameter Space} \label{appendix:Parameter Space}
As mentioned before, the likelihood calculation requires a full exploration of the relevant parameter space. This traversal of the parameter space was performed using a Markov chain Monte Carlo (MCMC) procedure, utilizing the \textsf{emcee} Python code package \cite{emcee}, which implements the Metropolis-Hastings method.
MCMC is a procedure for generating a random walk in the parameter space that, over time, draws a representative set of samples from the distribution, thus approximating the posterior PDF. Once the samples produced by MCMC are available, it is then possible to marginalize over parameters that are not of interest, by integrating over their possible values, which propagates the width of their distributions onto the result for the important parameters. 

\begin{figure}[t!]
\includegraphics[width=36em]{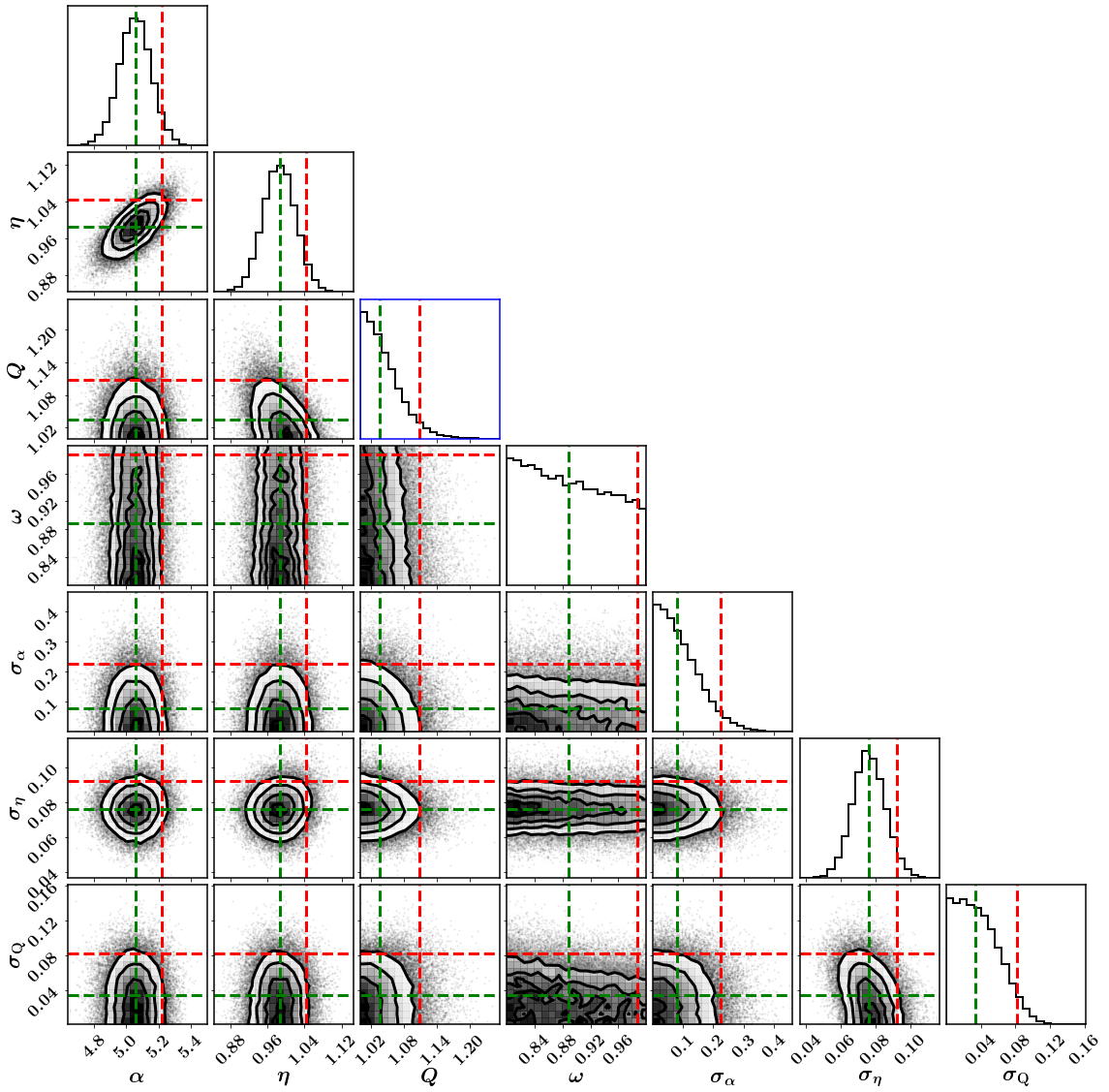}
\centering
\caption{\footnotesize Likelihood ratio corner plot resulting from MCMC analyzed on the $0.45<M_{\rm tot}/M_\odot < 1.05$ data segment. Main diagonal plots are 1D likelihood histograms after marginalization on all but one parameter; plots below the main diagonal are 2D likelihood histograms after marginalization on all but two parameters. Green dashed lines mark the median of each distribution; red dashed lines mark the 95th percentile. The blue frame highlights the $Q$ distribution, used to calculate the DM fractional mass upper bound.}
\label{fig:Results corner plot}
\end{figure}

For our use of MCMC, exploration of each of the parameters was initialized in some proximity to the maximum-likelihood point (which was calculated in the standard way), and then expanded from there to explore the parameter space, limited by very loose prior probability constraints. An example result for the $0.45<M_{\rm tot}/M_\odot < 1.05$ data segment is presented in Fig.~\ref{fig:Results corner plot}, in corner plot form. Likelihood ratio histograms are presented for each of the model parameters, and each pair of parameters, after marginalization on all but one and all but two parameters, respectively. The result for our DM quotient factor, $Q$, is highlighted with a blue frame.  
Similar figures were produced for each of the sample mass domains, resulting in 5 likelihood ratio histograms for $Q$ that were translated to the result presented in Fig.~\ref{fig:PLR results}. 

A feature of note in these results is the distribution of $\omega$, which for all mass domains appeared as close to flat, with a slight preference towards the lower values. This appears to result from a combination of $\omega$ being close to unity across the sample, along with lower values of $\omega$ allowing for a wider range of $Q$ and $\eta$ values to achieve a high likelihood score, creating a slight bias towards lower $\omega$ values in the MCMC search. 
Additionally, in all segments the $Q=1$,  $\sigma_Q=0$ values were not excluded, allowing for the reasonable possibility of no observable stellar DM density; it is entirely possible within our model for the stars to contain no DM whatsoever.

\end{document}